\documentclass[12pt]{article}
\usepackage{float}
\usepackage{graphics,epsfig}
\usepackage{bm}
\usepackage{slashed}
\usepackage{graphicx}
\usepackage{amsmath}
\usepackage{amsfonts}
\usepackage{amssymb}
\usepackage{color}
\usepackage{dcolumn}

\numberwithin{equation}{section}
\definecolor{darkgreen}{rgb}{0,0.35,0}

\begin{document}

\title{Black hole solution of Einstein-Born-Infeld-Yang-Mills theory}
\author{Kun Meng \thanks{mengkun@tjpu.edu.cn} ,  Da-Bao Yang\thanks{bobydbcn@163.com}\,, Zhan-Ning Hu\thanks{Zhanninghu@aliyun.com}\ \bigskip \\
{\small School of Science, Tianjin Polytechnic University,}\\
{\small Tianjin 300387, China}}
\maketitle

\begin{abstract}
A new four-dimensional black hole solution of Einstein-Born-Infeld-Yang-Mills theory is constructed, several degenerated forms of the black hole solution are presented. The related thermodynamical quantities are calculated, with which the first law of thermodynamics is checked to be satisfied. Identifying the cosmological constant as pressure of the system, the phase transition behaviors of the black hole in the extended phase space are studied.
\end{abstract}

\section{Introduction}
When considering electromagnetic field as the matter source to construct black hole solution in Einstein gravity, one usually use  the standard Maxwell's theory of the U(1) gauge field. However, when the field
is strong enough the linear Maxwell's theory becomes invalid, the non-linearities of electromagnetic field should be introduced as the correct electrodynamics. In 1930's, Born and Infeld proposed a non-linear electrodynamics with the aim of obtaining a finite value of the self-energy of electron\cite{BI}. Fradkin and Yeltsin reproduced Born-Infeld (BI) action in the framework of string theory\cite{Fradkin}. The D3-brane dynamics was also noticed to be governed by BI action\cite{Tseytlin}. Hoffmann first found a solution of Einstein gravity coupled to BI electromagnetic field\cite{Hoffmann}, which is devoid of essential singularity at the origin. Many black hole solutions of Einstein-Born-Infeld (EBI) theory with or without a cosmological constant have been found \cite{Oliveira,FernandoKrug,Dey,CaiBI}. Black hole solutions of Gauss-Bonnet and three order Lovelock gravity coupled to BI electromagnetic field have also been found in \cite{Wiltshire,DehghaniHendi}.

Since the condition for no-hair theorem in asymptotic (A)dS spacetime is much relaxed, besides the usually used matter sources such as scalar and Maxwell fields, it's natural to couple Yang-Mills (YM) field to gravity and construct black hole solutions. However, the YM equations are so complicated that early attempts of searching for black hole solutions of Einstein-Yang-Mills (EYM) theory were performed numerically. The first analytic black hole solution was found by Yasskin applying Wu-Yang ansatz\cite{Yasskin}. This kind of solutions have also been generalized to higher dimensions and higher derivative gravity\cite{Halilsoy1,Halilsoy2}. Regular non-minimal magnetic black hole solutions have been investigated recently in \cite{Balakin}.
To our knowledge, black hole solutions of Einstein gravity coupled to both BI electromagnetic field and YM field have not been studied, in this paper, we want to construct black hole solution of Einstein-Born-Infeld-Yang-Mills (EBIYM) gravity and study some properties of the black hole.

Among the  properties of black holes, thermodynamical properties have gained much attention in the past several decades. Since the early work\cite{HawkingPage} studying thermal phase transition between Schwarzchild-AdS black hole and AdS vacuum, the behaviors of phase transition of black holes have been studied intensively. In Refs.\cite{Myers1,Myers2}, the thermodynamics of charged AdS black hole have been studied and found there exist phase transitions between large/small black holes at the sacrifice of viewing the electric charge as intensive variable, the phase transition behavior resembles the one of van der Waals liquid-gas system. In order to solve the problem encountered in\cite{Myers1,Myers2}, Mann et al extended the phase space by identifying the cosmological constant as the pressure of the gravitational system\cite{Mann1,Mann}, then the thermodynamic volume conjugate to pressure can be defined, in this framework the phase transitions between large/small black holes are founded too. Subsequently, the study was extended to various other kinds of black holes\cite{CaiCao,Zhao1,Zhao2,Mo}. In this paper, we will also explore the phase transitions of the black hole given below in extended phase space.

The paper is organized as follows. In Section \ref{sectionsolution}, we present the black hole solution of four dimensional Einstein-BI-YM gravity and discuss its several degenerated forms. In Section \ref{sectionthermdynamics}, we calculate the thermodynamical quantities, check the first law of thermodynamics  and study the phase transitions of the black hole in extended phase space. Finally, some concluding remarks are given in Section \ref{sectionconclusion}.

\section{Black hole solution of Einstein-BI-YM theory\label{sectionsolution}}
We intend to search for a four dimensional black hole solution of EBIYM theory, the action of the theory is given by
\begin{align}
\mathcal{I}=\int d^4x \sqrt{-g}\left[ \frac{R-2\Lambda}{16\pi G}+L(F)-\textbf{Tr}\left(F^{(a)}_{\mu\nu}F^{(a)\mu\nu}\right)\right],\label{model}
\end{align}
where $G$ is the Newton constant, we take the convention $16\pi G=1$ in the following for convenience, $\textbf{Tr}(\cdot)$ means sum over the gauge group indices, and
\begin{align}
L(F)=4 \beta^2\left(1-\sqrt{1+\frac{F^{\mu\nu}F_{\mu\nu}}{2\beta^2}}\right).
\end{align}
Here $\beta$ is the BI parameter with dimension of mass. In the limit $\beta\rightarrow 0$, $L(F)\rightarrow0$. In the limit $\beta\rightarrow \infty$, $L(F)$ reduces to the standard Maxwell form
\begin{align}
L(F)=-F_{\mu\nu}F^{\mu\nu}+\mathcal{O}(F^4).
\end{align}

Take variation with respect to the metric, the Maxwell field $A_\mu$ and the YM field $A^{(a)}_\mu$ respectively, we have the equations of motion
\begin{align}
&G_{\mu\nu}+\Lambda g_{\mu\nu}=T^M_{\mu\nu}+T^{YM}_{\mu\nu},\label{metricEOM}\\
&\nabla_\mu\left(\frac{F^{\mu\nu}}{\sqrt{1+\frac{F_{\rho\sigma}F^{\rho\sigma}}{2\beta^2}}}\right)=0,\label{MaxEOM}\\
&\nabla_\mu F^{(a)\mu\nu}+f^{(a)}_{\;(b)(c)}A^{(b)}_\mu F^{(c)\mu\nu}=0,\label{YMEOM}
\end{align}
where $f^{(a)}_{\;(b)(c)}$ are the real structure constants of the gauge group and
\begin{align}
T^{EM}_{\mu\nu}&=\frac{1}{2}g_{\mu\nu}L(F)+\frac{2F_{\mu\rho}F_\nu^{\;\rho}}{\sqrt{1+\frac{F_{\rho\sigma}F^{\rho\sigma}}{2\beta^2}}},\nonumber\\
T^{YM}_{\mu\nu}&=-\frac{1}{2}g_{\mu\nu}F^{(a)}_{\rho\sigma}F^{(a)\rho\sigma}+2F^{(a)}_{\mu\rho}F_\nu^{(a)\rho}.\label{energymomentum}
\end{align}

In order to find a static solution of the field equations, we take a general metric ansatz
\begin{align}
ds^2=-f(r)dt^2+\frac{dr^2}{f(r)}+r^2d\Omega_{2}^2,\label{metricansatz}
\end{align}
where $d\Omega_{2}^2$ is the line element of 2-sphere.
Before solving eq.(\ref{metricEOM}), we need to solve equations of motion of the matter fields (\ref{MaxEOM}) and (\ref{YMEOM}) first. Under the electrostatic potential assumption, all other components of the strength tensor $F^{\mu\nu}$ vanish except $F^{tr}$, solving eq.(\ref{MaxEOM}) one obtains
\begin{align}
F^{tr}=\frac{\beta q}{\sqrt{\beta^2r^4+q^2}}.\label{Maxstrength}
\end{align}
where $q$ is an integration constant. Note that unlike the field strength in Maxwell electrodynamics, the one in BI electrodynamics is finite at $r=0$. To solve equations of motion of the YM field (\ref{YMEOM}), we take the magnetic Wu-Yang ansatz of the gauge potential\cite{Balakin}.
Since we want to find a black hole solution with global YM charge, the gauge group is supposed to be $SU(2)$, then the real structure constants $f_{(a)(b)(c)}$  are the complete antisymmetric symbols $\varepsilon_{(a)(b)(c)}$. We use the position dependent generators $\mathbf{t}_{(r)}, \mathbf{t}_{(\theta)}, \mathbf{t}_{(\varphi)}$ of the gauge group, the relations between the generators $\mathbf{t}_{(r)}, \mathbf{t}_{(\theta)}, \mathbf{t}_{(\varphi)}$ and the  standard $SU(2)$ generators are
\begin{align}
\mathbf{t}_{(r)}=\cos(\nu\varphi)\sin\theta \mathbf{t}_{(1)}+\sin(\nu\varphi)\sin\theta &\mathbf{t}_{(2)}+\cos\theta\mathbf{t}_{(3)},\nonumber\\
\mathbf{t}_{(\theta)}=\cos(\nu\varphi)\cos\theta \mathbf{t}_{(1)}+\sin(\nu\varphi)\cos\theta &\mathbf{t}_{(2)}-\sin\theta\mathbf{t}_{(3)},\nonumber\\
\mathbf{t}_{(\varphi)}=-\sin(\nu\varphi)\mathbf{t}_{(1)}+\cos(\nu\varphi)&\mathbf{t}_{(2)}.
\end{align}
where $\nu$ is a non-vanishing integer, as will become clear later, it relates to global YM charge. Using the commutation relations $
[\mathbf{t}_{(a)}, \mathbf{t}_{(b)}]=\varepsilon_{(a)(b)(c)}\mathbf{t}_{(c)}
$, where $a, b, c$ take values from 1 to 3 and $\varepsilon_{(1)(2)(3)}=1$, it is straightforward to check the following commutation relations
\begin{align}
[\mathbf{t}_{(r)}, \mathbf{t}_{(\theta)}]=\mathbf{t}_{(\varphi)},\;\;\;\;\;[\mathbf{t}_{(\theta)}, \mathbf{t}_{(\varphi)}]=\mathbf{t}_{(r)},\;\;\;\;\;
[\mathbf{t}_{(\varphi)}, \mathbf{t}_{(r)}]=\mathbf{t}_{(\theta)},
\end{align}
are satisfied. The gauge field characterized by Wu-Yang ansatz is of the form
\begin{align}
A_0^{(a)}=0,\;\;\;A_r^{(a)}=0,\;\;\;A_\theta^{(a)}=-\delta^{(a)}_{(\varphi)},\;\;\;A_\varphi^{(a)}=\nu \sin\theta\delta^{(a)}_{(\theta)},\label{YMchoice}
\end{align}
One can check that the equations of motion of YM field (\ref{YMEOM}) are satisfied under the choice of gauge (\ref{YMchoice}). Under the gauge (\ref{YMchoice}), the only non-vanishing component of the YM field strength tensor is
\begin{align}
\mathbf{F}_{\theta\varphi}=-\nu \sin\theta \mathbf{t}_{(r)}.\label{YMstrength}
\end{align}

With the field strength tensors (\ref{Maxstrength}), (\ref{YMstrength}) and the metric ansatz (\ref{metricansatz}), one can work out the expression of energy momentum tensor straightforwardly. Substituting eqs.(\ref{energymomentum}) and (\ref{metricansatz}) into eq.(\ref{metricEOM}) and solving eq.(\ref{metricEOM}) we find the following new black hole solution
\begin{align}
f(r)=1-\frac{m_0}{r}+\frac{1}{3}\left(2\beta^2-\Lambda\right) r^2+\frac{\nu^2}{r^2}-\frac{2\beta}{3}\sqrt{q^2+r^4\beta^2}+\frac{4q^2}{3r^2}\;_2F_1[\frac{1}{2}, \frac{1}{4}, \frac{5}{4}, -\frac{q^2}{r^4 \beta^2}].\label{solution}
\end{align}
Here $m_0$ is the mass parameter and $\;_2F_1[\frac{1}{2}, \frac{1}{4}, \frac{5}{4}, -\frac{q^2}{r^4 \beta^2}]$ is the Gaussian hypergeometric function. Before further analysis of the solution (\ref{solution}), we discuss its several limiting forms. When the parameter $\nu$ in (\ref{solution}) vanishes, the Einstein-BI black hole in (A)dS space is reproduced\cite{CaiBI}. When  $\beta\rightarrow\infty$,   BI electrodynamics reduces to the standard Maxwell form, the solution (\ref{solution}) reduces to
\begin{align}
f(r)=1-\frac{m_0}{r}-\frac{1}{3}\Lambda r^2+\frac{q^2}{r^2}+\frac{\nu^2}{r^2},\label{solbeta}
\end{align}
which is a new black hole with both electric charge and magnetic YM charge. When   $\beta\rightarrow0$, the electric charge vanishes automatically, one obtains a magnetic black hole with only YM charge
\begin{align}
f(r)=1-\frac{m_0}{r}-\frac{1}{3}\Lambda r^2+\frac{\nu^2}{r^2},\label{MegBH}
\end{align}
which is of the same form as the 4-dimensional (A)dS-Reinssner-Nordstr\"{o}m(RN) black hole. When $\nu=0$  and $\beta\rightarrow\infty$, the solution (\ref{solution}) degenerates to  RN black hole in (A)dS space.

\section{Thermodynamics\label{sectionthermdynamics}}
In this section, we will first calculate the thermodynamic quantities of the black hole obtained in the last section, then check the first law of thermodynamics, and finally extend the phase space to explore the phase transition behaviors of the black hole.
\subsection{Calculations of thermodynamical quantities}
Since the black hole (\ref{solution}) is asymptotic to (A)dS, in order to calculate the mass of the black hole, one should adopt the method proposed by Abbott and Deser\cite{AD}, which gives
\begin{align}
M=8\pi m_0.
\end{align}
Note that the area of unit 2-sphere $\omega_2=4\pi$ is used. In order to calculating the temperature, one should take the Euclidean continuation
($t\rightarrow-i\tau$) and demand the absence of conical singularity at the horizon, the period of the Euclidean time $\tau$ is the inverse of Hawking temperature $1/T$, one obtains
\begin{align}
T=\frac{1}{4\pi r_+}\left(1+2\beta^2r_+^2-\Lambda r_+^2-\frac{\nu^2}{r_+^2}-2\beta\sqrt{q^2+r_+^4\beta^2}\right),
\end{align}
where $r_+$ is the radius of the outmost horizon of the black hole. In the model (\ref{model}) we considered, entropy of the black hole (\ref{solution}) obeys the area law, which is
\begin{align}
S=16\pi^2 r_+^2,
\end{align}
note that the convention $16\pi G=1$ is adopted. The electric charge is defined as
\begin{align}
Q_e=\frac{1}{4\pi}\int \ast F d\Omega^2=q,\label{electriccharge}
\end{align}
where $\ast F$ represents Hodge dual of the strength tensor and the integration is performed on the $t=const$ and $r\rightarrow\infty$ hypersurface. The electrostatic potential $\Phi$, measured at infinity with respect to the horizon, is defined as
\begin{align}
\Phi=A_\mu \chi^\mu|_{r\rightarrow\infty}-A_\mu \chi^\mu|_{r=r_+}.
\end{align}
Our calculation gives $\Phi=\frac{q}{r_+}\;_2F_1[\frac{1}{2}, \frac{1}{4}, \frac{5}{4}, -\frac{q^2}{r_+^4 \beta^2}]$. Since the gauge group is $SU(2)$,  we can define a global YM charge, for which we take the definition\cite{Corichi,Kleihaus,Lu}
\begin{align}
Q_{_{YM}}=\frac{1}{4\pi}\int\sqrt{F^{(a)}_{\theta\varphi}F^{(a)}_{\theta\varphi}}d\theta d\varphi.\label{YMcharge}
\end{align}
Just as the definition of electric charge, the integration in (\ref{YMcharge}) is performed on the $t=const$ and $r\rightarrow\infty$ hypersurface too. Our calculation gives
\begin{align}
Q_{_{YM}}=-\nu.
\end{align}
In order to preserve the YM charge being positive, $\nu$ is taken to be negative. Now with all the thermodynamic quantities above in hand, one can check that the first law of thermodynamics
\begin{align}
dM=TdS+\Phi dQ_e+UdQ_{_{YM}}
\end{align}
is satisfied. Where $U$ defined by $U\equiv\left(\frac{\partial M}{\partial Q_{_{YM}}}\right)_{S,Q_e}$ is the thermodynamic potential conjugate to the YM charge $Q_{_{YM}}$.
\subsection{Phase Transition in extended phase space}
In extended phase space, cosmological constant is identified as  pressure of the gravitational system
\begin{align}
P=-\frac{\Lambda}{8\pi}.
\end{align}
There are some reasons to do so. First, one can suppose there exist  more fundamental theories, where the cosmological constant arises as vacuum expectation value. Second, the Smarr relation becomes inconsistent with the first law of thermodynamics unless the variation of $\Lambda$ is included\cite{Mann1}. Now the mass of the black hole is viewed as enthalpy $H\equiv M$ rather than internal energy of the system\cite{Kastor}, then the Gibbs free energy of the system is  $G=H-TS$.
If the BI parameter $\beta$ is viewed as a free thermodynamic variable too, one can check the first law of thermodynamics in extended phase space
\begin{align}
dH=TdS+\Phi dQ_e+UdQ_{_{YM}}+VdP+\mathcal{B}d\beta
\end{align}
is satisfied, where
\begin{align}
V=\left(\frac{\partial H}{\partial P}\right)_{S,\beta,Q_e,Q_{_{YM}}}=\frac{64\pi^2   r_+^3}{3}\label{volume}
\end{align}
is the thermodynamic volume conjugate to the pressure, and
\begin{align}
\mathcal{B}=\left(\frac{\partial H}{\partial \beta}\right)_{S,P,Q_e,Q_{_{YM}}}=\frac{4\pi}{3}\left(8\beta r_+^3-8r_+\sqrt{r_+^4\beta^2+Q_e^2}+\frac{4 Q_e^2}{\beta r_+}\;_2F_1[\frac{1}{2}, \frac{1}{4}, \frac{5}{4}, -\frac{Q_e^2}{r_+^4 \beta^2}]\right).
\end{align}
is the thermodynamic quantity conjugate to $\beta$.

Before studying the phase transitions of the black hole (\ref{solution}) for a general $\beta$, let's first discuss two limiting cases $\beta\rightarrow0$ and $\beta\rightarrow\infty$. When $\beta\rightarrow0$, in terms of the temperature and horizon radius, the pressure of the system can be written as
\begin{align}
P=\frac{T}{2r_+}-\frac{1}{8\pi r_+^2}+\frac{Q_{_{YM}}^2}{8\pi r_+^4}.\label{pressurebeta0}
\end{align}
The critical points of the thermal phase transition is determined by the following critical equations
\begin{align}
\frac{\partial P}{\partial r_+}\bigg|_{r_+=r_c,T=T_c}=\frac{\partial^2 P}{\partial r_+^2}\bigg|_{r_+=r_c,T=T_c}=0.\label{criticaleq}
\end{align}
The critical equations (\ref{criticaleq}) can be solved analytically, which gives the critical horizon radius
\begin{align}
r_{c}=\sqrt{6}Q_{_{YM}},
\end{align}
and the critical temperature
\begin{align}
T_c=\frac{1}{3\sqrt{6}\pi Q_{_{YM}}}.
\end{align}
It's easy to see that the critical horizon radius and temperature and consequently the critical pressure are solely determined by the YM charge. This is because, as previously mentioned, when $\beta\rightarrow0$ the electric charge vanishes automatically and the black hole carries only YM charge as shown in eq.(\ref{MegBH})

\begin{figure}[h]
\begin{center}
\includegraphics[width=.45\textwidth]{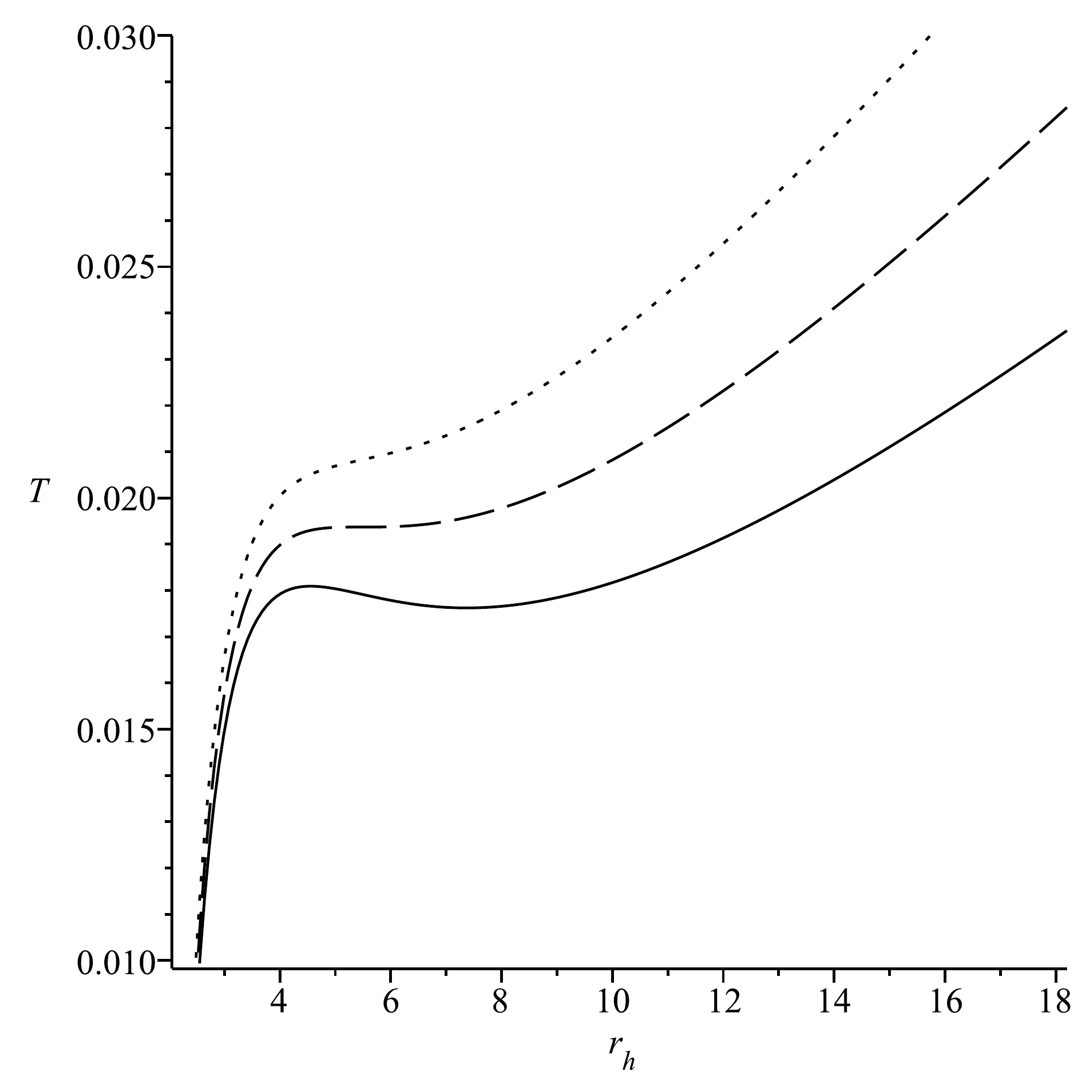}
\includegraphics[width=.45\textwidth]{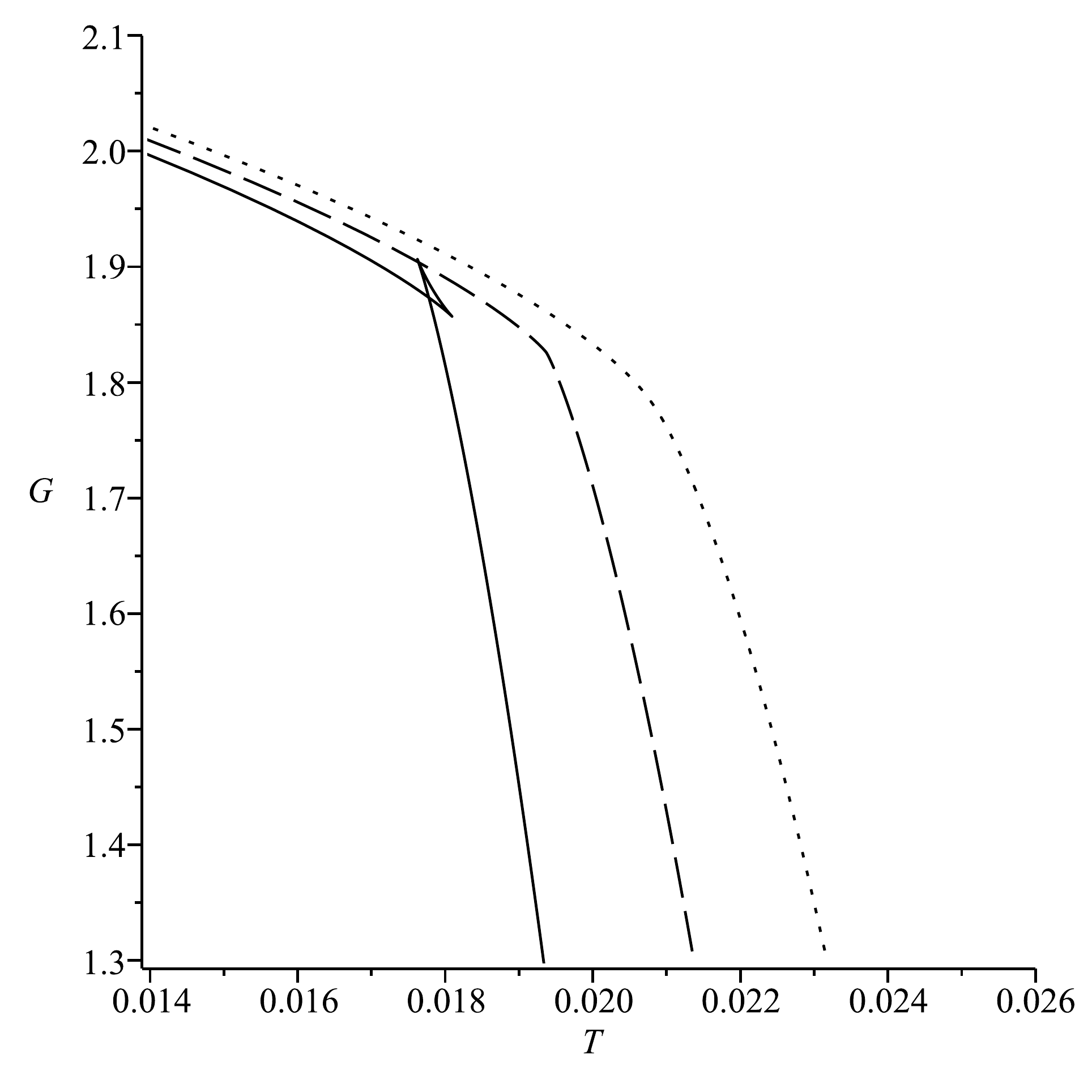}
\end{center}
\caption{Isobaric plots for $Q_{_{YM}}=2, Q_e=1$. The isobars on the $T-r_+$ plot correspond one-to-one with the $G-T$ plot. On both the left and the right plots, from top to bottom, the pressures are respectively $P=1.2P_c$(dotted), $P=P_c$(dashed) and $P=0.8P_c$(solid).}
\label{fig1}
\end{figure}

When $\beta\rightarrow\infty$, this case corresponds to the black hole described by the metric (\ref{metricansatz}) with $f(r)$ given in (\ref{solbeta}). The pressure of the system is given by
\begin{align}
P=\frac{T}{2r_+}-\frac{1}{8\pi r_+^2}+\frac{Q_e^2+Q_{_{YM}}^2}{8\pi r_+^4}.\label{pressure2}
\end{align}
Solving the critical equations (\ref{criticaleq}), we obtain the critical horizon radius
\begin{align}
r_{c}=\sqrt{6(Q_e^2+Q_{_{YM}}^2)},
\end{align}
and the critical temperature
\begin{align}
T_c=\frac{1}{3\pi \sqrt{6(Q_e^2+Q_{_{YM}}^2)}}.
\end{align}
The critical horizon radius, temperature and pressure are determined by both electric and YM charge.
Now, for the black hole (\ref{solbeta}) by taking proper values of the related parameters we can plot the isobars as displayed in Fig.\ref{fig1}. On the $T-r_+$ plot of Fig.\ref{fig1}, the dotted line is the isobar of $P>P_c$, in this case there is only a single phase which is in analogy to the thermal behavior of  ``ideal gas''. The dashed line is the isobar of $P=P_c$. The solid line is the isobar of $P<P_c$, one can see that there are two branches of black holes whose temperatures increase as the increase of horizon radius, one branch is in small radius region, the other one is in large radius region. Between the two branches, there is a region whose temperature decreases as the increase of horizon radius, which means the thermal expansion coefficient is negative, this region corresponds to an unstable phase. For appropriate value of temperature, two horizon radius are permitted, thus large/small black holes phase transition occurs. This kind of thermal behavior resembles the one of  van der Waals gas/liquid system. The phase transition can also be seen in the $G-T$ plot, where ``swallow tail'' appears when $P<P_c$, just as the solid line shows us. The dashed line on the $G-T$ plot is the isobar of $P=P_c$, and the dotted line corresponding to $P>P_c$ describes ``ideal gas'' phase.

\begin{table}
\centering
\begin{tabular}{cccccc}
\hline\hline
$\beta$ & $Q_{_{YM}}$ &$r_{c}$ & $T_c$ & $P_c$ & $P_c r_{c}/T_c$ \\ [0.5ex]
\hline
10 & 3 & 7.7459654& 0.0136979 & 0.0003316 & 0.1875150\\
\hline
10 & 2 & 5.4772185 & 0.0193717 & 0.0006631 & 0.1875024\\
\hline
10 & 1 &3.4640315 & 0.0306296 & 0.0016579 &0.1874989\\
\hline
1 & 2 &5.4765161 & 0.0193728 & 0.0006632 &0.1874807\\
\hline
0.1 & 2 &5.4113483 & 0.0194752 & 0.0006714 &0.1865541\\
\hline
\end{tabular}
\label{table}
\caption{Critical values for $Q_e=1$.}
\end{table}

\begin{figure}[h]
\begin{center}
\includegraphics[width=.45\textwidth]{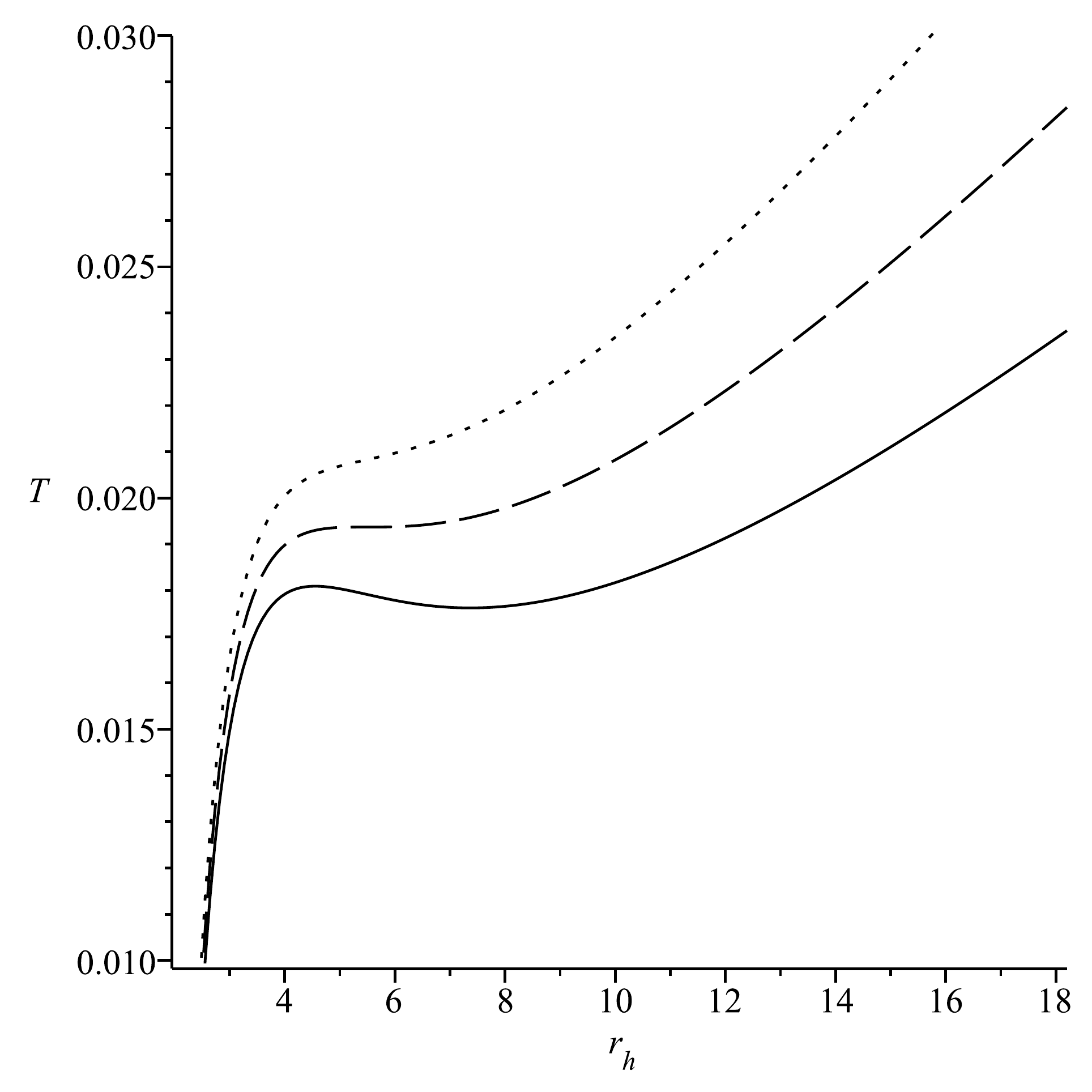}
\includegraphics[width=.45\textwidth]{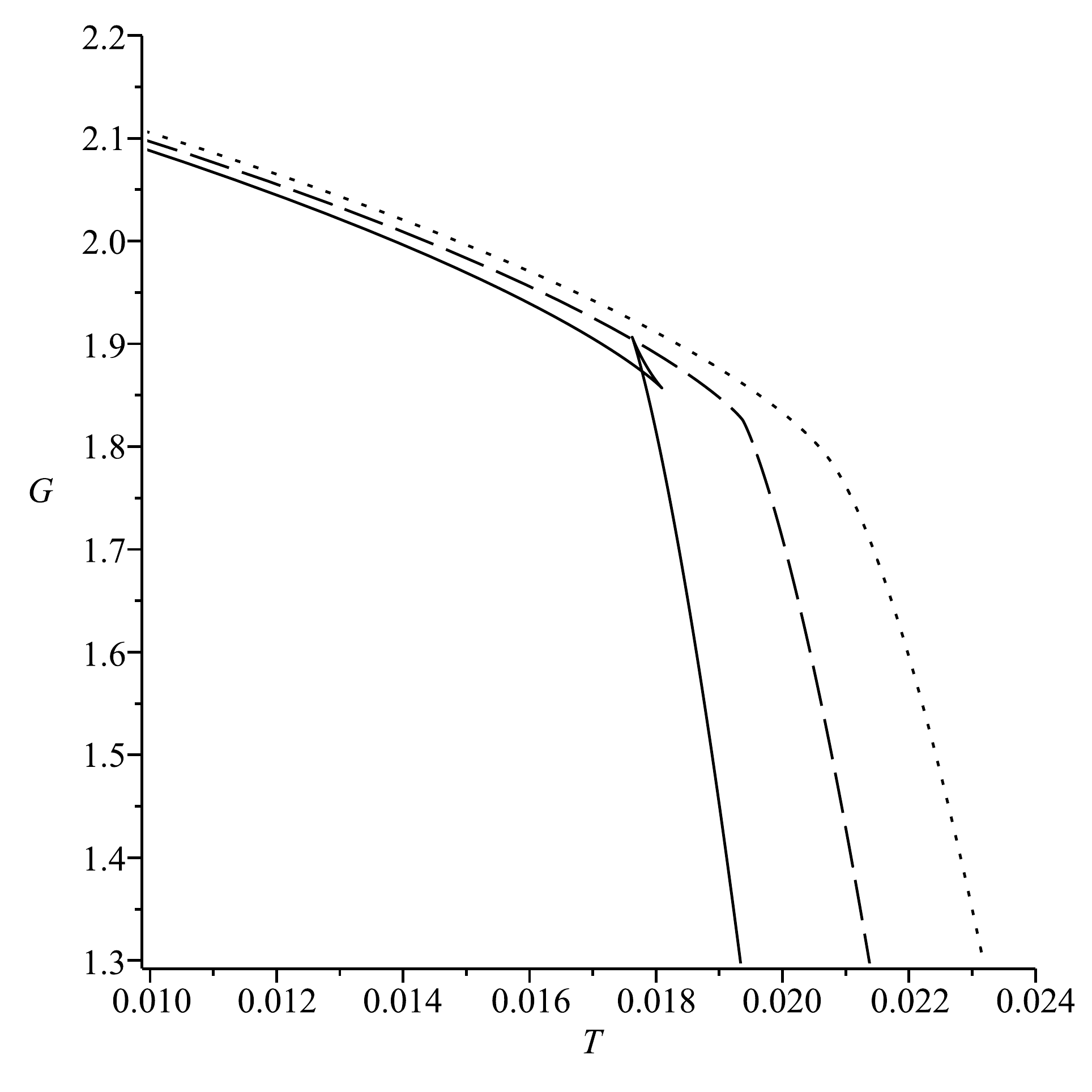}
\end{center}
\caption{Isobaric plots for $Q_{_{YM}}=2, Q_e=1, \beta=10$. The isobars on the $T-r_+$ plot and the $G-T$ plot are in one-to-one correspondence  with each other. On both the left and the right plots, from top to bottom, the pressures are respectively $P=1.2P_c$(dotted), $P=P_c$(dashed) and $P=0.8P_c$(solid).}
\label{fig2}
\end{figure}

\begin{figure}[h]
\begin{center}
\includegraphics[width=.45\textwidth]{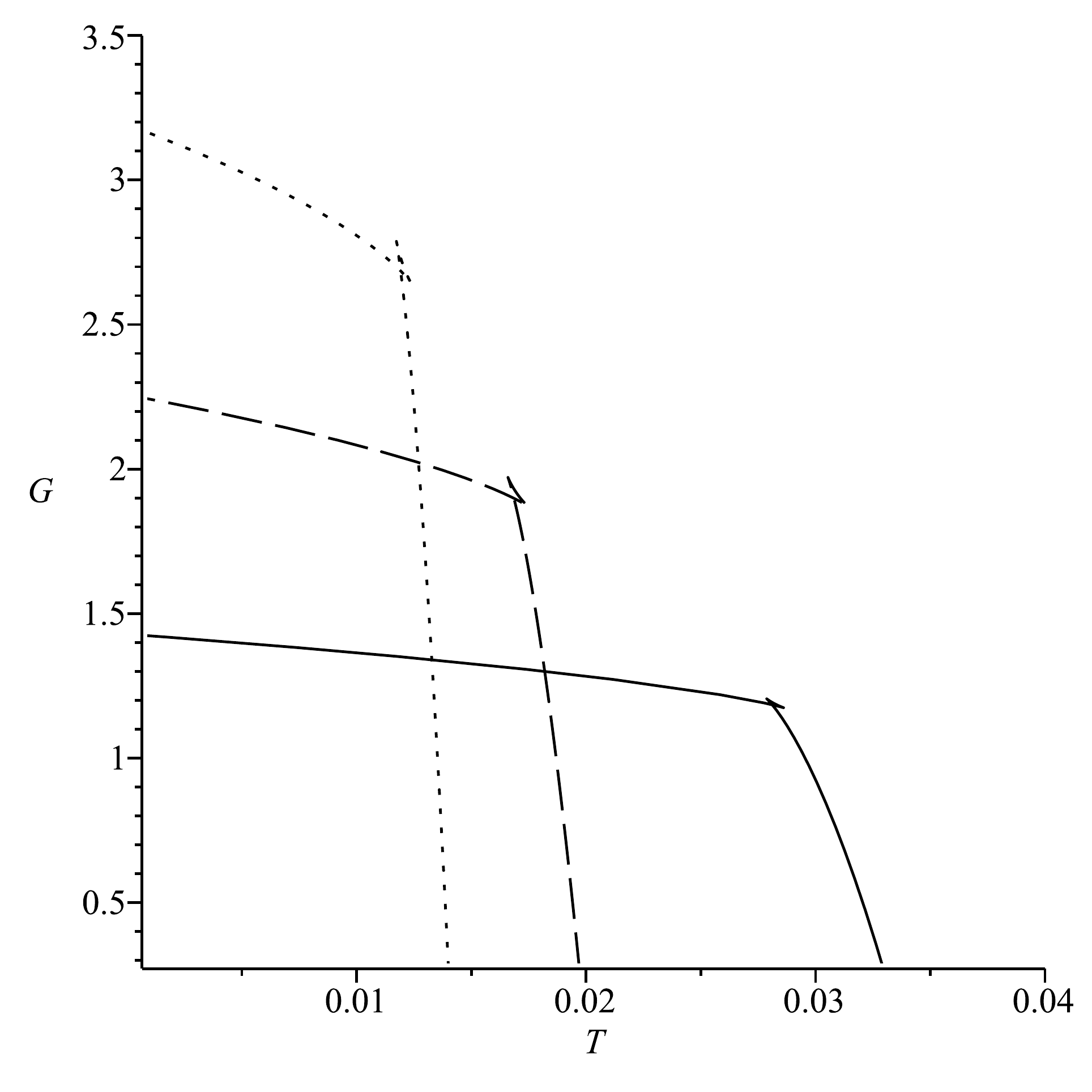}
\includegraphics[width=.45\textwidth]{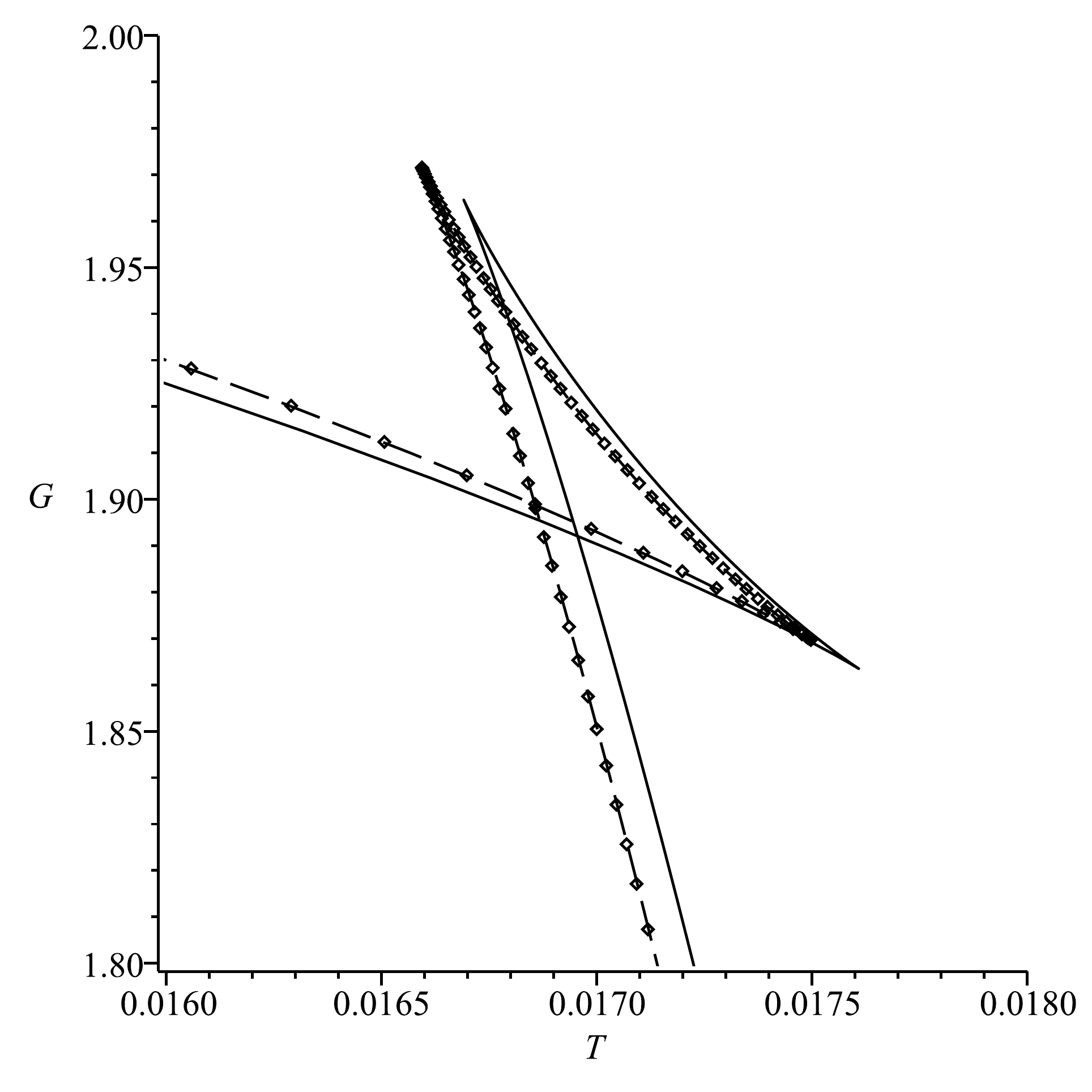}
\end{center}
\caption{Isobaric plots that reflect the effects of YM charge(left plot) and BI parameter(right plot). On the left plot, the parameters of all the isobars are taken as $Q_e=1, \beta=10$ and $P=0.7P_c$, the difference is, the solid, dashed and dotted lines correspond to $Q_{_{YM}}=1, 2, 3$ respectively. On the right plot, the parameters of all the isobars are taken as $Q_e=1, Q_{_{YM}}=2$ and $P=0.7P_c$,  the solid, dotted and dashed lines correspond to $\beta=0.1, 1, 10$ respectively.}
\label{fig3}
\end{figure}

Now, let's consider phase transition of the black hole (\ref{solution}) for a general value of $\beta$. For general $\beta$, the pressure of the system is
\begin{align}
P=-\frac{1}{8\pi r_+^2}+\frac{T}{2r_+}+\frac{\beta^2}{4\pi}\sqrt{1+\frac{Q_e^2}{\beta^2r_+^4}}-\frac{\beta^2}{4\pi}+\frac{Q_{_{YM}}^2}{8\pi r_+^4}.
\end{align}
In this case, the critical equations (\ref{criticaleq}) are too complicated to be solved analytically, we solve them numerically, the results are collected in the table. From the data in the table, one learns that  with the increase of $\beta$ or $Q_{_{YM}}$, $T_c$ and $P_c$ decrease while $r_c$ and the ratio $\frac{P_c r_{c}}{T_c}$ increase.
Fig.\ref{fig2} gives the isobaric plots near the  critical point of $P_c=0.0006631$ and Fig.\ref{fig3} displays the effects of YM charge and BI parameter on the critical pressures.
On the $T-r_+$ plot of Fig.\ref{fig2}, just as the case of $\beta\rightarrow\infty$, the $P>P_c$ isobar describes the ``ideal gas'' phase. The isobar of $P<P_c$ shows that the ``large black hole'' region and ``small black hole'' region are thermodynamical stable while the medium region is thermodynamical unstable, which indicates the existence of phase transition. On the $G-T$ plot of Fig.\ref{fig2}, the isobars show that when $P<P_c$ phase transition characterized by the``swallow tail'' occurs.
On the right plot of Fig.\ref{fig3} one can see that, although corresponding to different orders of magnitude of $\beta$, the dotted line($\beta=1$) and the dashed line($\beta=10$) are almost coincident, and the deviation of the solid line($\beta=0.1$) from the dotted or dashed one is small, it seems that the influence of the values of $\beta$ to critical pressure is not that large just as the authors found in\cite{Mo}. This may be attributed to the parameter region we choose.

\section{Conclusion\label{sectionconclusion}}
In this paper, we construct a  black hole solution of Einstein gravity coupled to BI electromagnetic field and SU(2) YM field. When the parameter $\nu=0$, one reproduces the Einstein-BI balck hole in (A)dS space. When the BI parameter $\beta\rightarrow\infty$, the solution degenerates to a new black hole with electrostatic charge and magnetic YM charge. When  $\beta\rightarrow 0$, the solution reduces to a black hole with only YM charge. When the parameters $\nu=0$ and $\beta\rightarrow\infty$, one obtains the (A)dS-RN black hole. We calculate the thermodynamical quantities such as mass, temperature, entropy, electric charge and YM charge etc., and check that the fist law of thermodynamics is satisfied.

We study the phase transition behaviors of the black hole in extended phase space by identifying the cosmological constant as pressure of the system. The critical equations can be solved analytically when taking the $\beta\rightarrow\infty$ limit while they have to be solved numerically for general $\beta$. After plotting the isobars one sees that, when above the critical pressure there is only an ``ideal gas'' phase, when below the critical pressure there are two stable region corresponding respectively to large/small black hole phases with an unstable medium region, which implies the existence of the large/small black hole phase transition, this thermal behavior of black hole is in analogy to the one of van der Waals gas. The effects of $\beta$ and $Q_{_{YM}}$ on critical pressures are also considered and displayed in the isobaric plots.

\section*{Acknowledgment}
KM would like to thank Professor Liu Zhao for valuable discussions. This work is supported by the National Natural Science Foundation of China (NSFC) under the
grant numbers 11447153 (for KM) and 11447196 (for DBY).

\providecommand{\href}[2]{#2}\begingroup
\footnotesize\itemsep=0pt
\providecommand{\eprint}[2][]{\href{http://arxiv.org/abs/#2}{arXiv:#2}}

\bibliographystyle{utcaps}
\bibliography{papers1,papers2,books}

\begin{thebibliography}{}
\bibitem{BI}
M. Born and L. Infeld, ``Foundations of the New Field Theory,''  Proc. Roy. Soc. Lond. A $\mathbf{144}$ (1934) 425.

\bibitem{Fradkin}
E. S. Fradkin and  A.A. Tseytlin , ``Non-linear electrodynamics from quantized strings,''  Phys. lett. B $\mathbf{163}$ (1985) 123.

\bibitem{Tseytlin}
A. A. Tseytlin, ``Vector field effective action in the open superstring theory,''  Nucl. Phys. B $\mathbf{276}$ (1986) 391.

\bibitem{Hoffmann}
B. Hoffmann, ``Gravitational and Electromagnetic Mass in the BI Electrodynamics,''  Phys. Rev.  $\mathbf{47}$ (1935) 877.

\bibitem{Oliveira}
H P de Oliveira, ``Non-linear charged black holes,''  Class. Quantum Grav.  $\mathbf{11}$ (1994) 1469.

\bibitem{FernandoKrug}
S. Fernando and D. Krug, ``Charged Black Hole Solutions in Einstein-BI gravity with a Cosmological constant,''  Gen. Rel. Grav. $\mathbf{35}$ (2003)129-137 [\eprint{hep-th/0306120}].

\bibitem{Dey}
Tanay Kr. Dey, ``BI black holes in the presence of a cosmological constant,''  Phys.Lett. B  $\mathbf{595}$ (2004)484-490 [\eprint{hep-th/0406169}].

\bibitem{CaiBI}
Rong-Gen Cai, Da-Wei Pang, Anzhong Wang, ``BI Black Holes in (A)dS Spaces,''  Phys.Rev.D $\mathbf{70}$ (2004)124034 [\eprint{hep-th/0410158}].

\bibitem{Wiltshire}
D.L. Wiltshire, ``Spherically symmetric solutions of Einstein-Maxwell theory with a Gauss-Bonnet term,''  Phys.Lett. B  $\mathbf{169}$ (1986)36.

\bibitem{DehghaniHendi}
M. H. Dehghani, N. Alinejadi, S. H. Hendi, ``Topological Black Holes in Lovelock-BI Gravity,''  Phys.Rev.D $\mathbf{77}$ (2008)104025 [\eprint{0802.2637}].

\bibitem{Yasskin}
Philip B. Yasskin, ``Solutions for gravity coupled to massless gauge fields,''  Phys.Rev.D $\mathbf{12}$ (1975)2212.

\bibitem{Halilsoy1}
S. H. Mazharimousavi and M. Halilsoy, ``5D-Black Hole Solution in Einstein-YM-Gauss-Bonnet Theory,''  Phys.Rev.D $\mathbf{76}$ (2007)087501 [\eprint{0801.1562}].


\bibitem{Halilsoy2}
S. H. Mazharimousavi and M. Halilsoy, ``Black Hole solutions in Einstein-Maxwell-YM-Gauss-Bonnet Theory,''  J. Cosmol. Astropart. Phys. $\mathbf{12}$ (2008)005 [\eprint{0801.2110}].

\bibitem{Balakin}
Alexander B. Balakin, Jos\'{e}P. S. Lemos, Alexei E. Zayats, ``Regular nonminimal magnetic black holes in spacetimes with a cosmological constant,''   Phys.Rev.D $\mathbf{93}$ (2016)024008 [\eprint{1512.02653}].

\bibitem{HawkingPage}
S. Hawking and D. N. Page, ``Thermodynamics of Black Holes in anti-De Sitter Space,''  Commun.Math.Phys.  $\mathbf{87}$ (1983) 577.

\bibitem{Myers1}
Andrew Chamblin, Roberto Emparan, Clifford V. Johnson, Robert C. Myers, ``Charged AdS black holes and catastrophic holography,''   Phys.Rev. D $\mathbf{60}$ (1999) 064018 [\eprint{hep-th/9902170}].


\bibitem{Myers2}
Andrew Chamblin, Roberto Emparan, Clifford V. Johnson, Robert C. Myers, ``Holography, thermodynamics and fluctuations of charged AdS black holes ,''   Phys.Rev. D $\mathbf{60}$ (1999) 104026 [\eprint{hep-th/9904197}].

\bibitem{Mann1}
David Kubiznak, Robert B. Mann, ``P-V criticality of charged AdS black holes,''   JHEP $\mathbf{1207}$ (2012)033 [\eprint{1205.0559}].


\bibitem{Mann}
Sharmila Gunasekaran, David Kubiznak, Robert B. Mann, ``Extended phase space thermodynamics for charged and rotating black holes and BI vacuum polarization,''   JHEP $\mathbf{1211}$ (2012)110 [\eprint{1208.6251}].

\bibitem{CaiCao}
Rong-Gen Cai, Li-Ming Cao, Li Li, Run-Qiu Yang, ``P-V criticality in the extended phase space of Gauss-Bonnet black holes in AdS space,''   JHEP $\mathbf{1309}$ (2013)005 [\eprint{1306.6233}].

\bibitem{Zhao1}
Wei Xu, Hao Xu, Liu Zhao, ``Gauss-Bonnet coupling constant as a free thermodynamical variable and the associated criticality,''   Eur. Phys. J. C $\mathbf{74}$ (2014)2970 [\eprint{1311.3053}].

\bibitem{Zhao2}
Hao Xu, Wei Xu, Liu Zhao, ``Extended phase space thermodynamics for third order Lovelock black holes in diverse dimensions,''   Eur. Phys. J. C $\mathbf{74}$ (2014)3074 [\eprint{1405.4143}].

\bibitem{Mo}
Jie-Xiong Mo and  Wen-Biao Liu, ``P-V  criticality of topological black holes in Lovelock-Born-Infeld gravity,''   Eur. Phys. J. C $\mathbf{74}$ (2014)2836 [\eprint{1401.0785}].


\bibitem{AD}
L. F. Abbott, Stanley Deser, ``Stability of Gravity with a Cosmological Constant,''  Nucl. Phys. B $\mathbf{195}$ (1982) 76-96.

\bibitem{Corichi}
A. Corichi, U. Nucamendi, D. Sudarsky, ``Einstein-YM Isolated Horizons: Phase Space, Mechanics, Hair and Conjectures,''   Phys.Rev.D $\mathbf{62}$ (2000)044046 [\eprint{gr-qc/0002078}].

\bibitem{Kleihaus}
B. Kleihaus, J. Kunz, F. Navarro-Lerida, ``Rotating Einstein-YM Black Holes,''   Phys.Rev.D $\mathbf{66}$ (2002)104001 [\eprint{gr-qc/0207042}].

\bibitem{Lu}
Zhong-Ying Fan, H. L\"{u}, ``SU(2)-Colored (A)dS Black Holes in Conformal Gravity,''   JHEP $\mathbf{1502}$ (2015)013 [\eprint{1411.5372}].

\bibitem{Kastor}
David Kastor, Sourya Ray, Jennie Traschen, ``Enthalpy and the Mechanics of AdS Black Holes,''   Class.Quant.Grav. $\mathbf{26}$ (2009)195011 [\eprint{0904.2765}].




\end{thebibliography}
\end{document}